# The minimal principals of Hermitian matrices and the negativity of bipartite of qubit states

João Luzeilton de Oliveira, Rubens Viana Ramos

*Departamento de Engenharia de Teleinformática – Universidade Federal do Ceará - DETI/UFC*
*C.P. 6007 – Campus do Pici - 60755-640 Fortaleza-Ce Brasil*

Quantum entanglement is an enigmatic and powerful property that has attracted much attention due to its usefulness in new ways of communications, like quantum teleportation and quantum key distribution. Much effort has been done to quantify entanglement. Indeed, there exist some well-established separability criterion and analytical formulas for the entanglement of bipartite systems. In some of these, the crucial elements are the eigenvalues of the partial transpose of the density matrix. In this paper, we show that one can also have information about the entanglement of bipartite state, in $C^2 \otimes C^2$, looking at the minimal principals of the partial transpose.

## 1. Introduction

Entanglement is the physical property fundamental for development of quantum information processing such as quantum cryptographic protocols and quantum computing. On the other hand, the density matrix contains all information available on all possible future developments of a quantum state. Indeed, the separability (if a quantum state is entangled or not) as well as the quantification of the entanglement of bipartite quantum systems can be obtained using the density matrix through its partial transpose [1-7]. In particular, the important property of the partial transpose of the density matrix is the set of eigenvalues. In this article, we propose a different approach in the treatment of the density matrix in order to obtain information about the entanglement of $C^2 \otimes C^2$ states. Instead of using the eigenvalues of the partial transpose, here we use the minimal principals in order to state a separability criterion and an entanglement measure. The minimal principals have traditionally been used in order to determine the local stability of non-linear systems [8]. Although in larger number than eigenvalues, their advantage is the easiness of calculation.

This work is outlined as follows. In Section 2 the separability criterion and entanglement measure for bipartite of qubit states based on the eigenvalues of the partial transpose are reviewed. In Section 3 the separability criterion and entanglement measure based on the minimal principals are provided. At last, the conclusions are presented in Section 4.

## 2. Separability criterion and entanglement measure based on eigenvalues

The separability criterion for 2x2 and 2x3 systems was proposed by Peres and the Horodecki family in [1,2]. It states that if the partial transpose of the density matrix is positive semi-definite, that is, if it does not have negatives eigenvalues, then the state is separable otherwise it is entangled. Explicitly, let $\Gamma$ be the density matrix of a bipartite system with density matrices of the individual parts $\rho_a = Tr_b(\Gamma)$ and $\rho_b = Tr_a(\Gamma)$, where

$Tr_{a(b)}$ stands for the partial trace with respect to subsystem $a(b)$. The elements of $\Gamma$ can be given as:

$$\Gamma_{m\mu,n\nu} = \langle A_m, B_\mu | \Gamma | A_n, B_\nu \rangle, \tag{1}$$

where $\{A_m\}$ and $\{B_\mu\}$ are (arbitrary) orthonormal basis, in the Hilbert space, for the individual systems, $A$ and $B$, respectively. Using this representation, the partial transpose relative to the $A$ system is given by [1,2]:

$$\Gamma^{T_A}_{m\mu,n\nu} = \Gamma_{n\mu,m\nu}. \tag{2}$$

If $\Gamma$ is separable (disentangled) it has the form:

$$\Gamma = \sum_i p_i \rho_a^i \otimes \rho_b^i = \sum_i p_i (\rho_a^i)_{mn} (\rho_b^i)_{\mu\nu} \tag{3}$$

$$\sum_i p_i = 1 \tag{4}$$

In this case, Eq. (2) is equivalent to:

$$\Gamma^{T_A} = \sum_i p_i (\rho_a^i)^T \otimes \rho_b^i. \tag{5}$$

The Peres-Horodecki criterion is a binary condition: The partial transpose has or has not any negative eigenvalue? A quantitative version of this criterion was proposed by Vidal and Werner by using a new quantity, based on the trace norm of the partial transpose, called negativity [3]:

$$N_e(\Gamma) \equiv \frac{\|\Gamma^{T_A}\|_1 - 1}{2} \tag{6}$$

$$\|\Gamma^{T_A}\|_1 = Tr\left(\sqrt{(\Gamma^{T_A})^+ \Gamma^{T_A}}\right) \tag{7}$$

The negativity is an entanglement monotone [7] and, hence, it can be used as a measure of entanglement. It is null when the state is separable. At last, the negativity can also be written as the absolute value of the sum of the negative eigenvalues of $\Gamma^{T_A}$.

## 3. Minimal Principals of Hermitian Matrices

The positive semi-definiteness of a Hermitean matrix can be checked through Sylvester's theorem: A Hermitian matrix $\rho$ is positive semi-definite if all minimal principals of $\rho$ are not negatives. The minimal principals of a 4x4 Hermitian matrix are:

$$m_1^1 = \rho_{11}, \ m_2^1 = \rho_{22}, \ m_3^1 = \rho_{33}, \ m_4^1 = \rho_{44} \tag{8}$$

$$m_1^2 = \det\begin{bmatrix} \rho_{11} & \rho_{12} \\ \rho_{21} & \rho_{22} \end{bmatrix}, \ m_2^2 = \det\begin{bmatrix} \rho_{11} & \rho_{13} \\ \rho_{31} & \rho_{33} \end{bmatrix}, \ m_3^2 = \det\begin{bmatrix} \rho_{11} & \rho_{14} \\ \rho_{41} & \rho_{44} \end{bmatrix}, \ m_4^2 = \det\begin{bmatrix} \rho_{22} & \rho_{23} \\ \rho_{32} & \rho_{33} \end{bmatrix},$$

$$m_5^2 = \det\begin{bmatrix} \rho_{22} & \rho_{24} \\ \rho_{42} & \rho_{44} \end{bmatrix}, \ m_6^2 = \det\begin{bmatrix} \rho_{33} & \rho_{34} \\ \rho_{43} & \rho_{44} \end{bmatrix} \tag{9}$$

$$m_1^3 = \det\begin{bmatrix} \rho_{11} & \rho_{12} & \rho_{13} \\ \rho_{21} & \rho_{22} & \rho_{23} \\ \rho_{31} & \rho_{32} & \rho_{33} \end{bmatrix}, \ m_2^3 = \det\begin{bmatrix} \rho_{11} & \rho_{12} & \rho_{14} \\ \rho_{21} & \rho_{22} & \rho_{24} \\ \rho_{41} & \rho_{42} & \rho_{44} \end{bmatrix}, \ m_3^3 = \det\begin{bmatrix} \rho_{11} & \rho_{13} & \rho_{14} \\ \rho_{31} & \rho_{33} & \rho_{34} \\ \rho_{41} & \rho_{43} & \rho_{44} \end{bmatrix},$$

$$m_4^3 = \det\begin{bmatrix} \rho_{22} & \rho_{23} & \rho_{24} \\ \rho_{32} & \rho_{33} & \rho_{34} \\ \rho_{42} & \rho_{43} & \rho_{44} \end{bmatrix} \tag{10}$$

$$m_1^4 = \det(\rho) \tag{11}$$

where $\rho_{ij}$ are the elements of the Hermitian matrix $\rho$ and $m_l^k$ is the $l$-th minimal principal of order $k$.

Comparing the Peres-Horodecki separability criterion and the Sylvester's theorem, one can rewrite Peres-Horodecki separability criterion as: *if the partial transpose of the density matrix does not have negatives minimal principals then the state is separable otherwise it is entangled*. However, differently from the idea of negativity, the absolute value of the sum of the negatives minimal principals is not an entanglement measure.

In order to state an entanglement measure based on the minimal principals let us use the following polynomial:

$$x^4 - S_1 x^3 + S_2 x^2 - S_3 x + S_4 = 0 \tag{12}$$

$$S_i = \sum_n m_n^i \tag{13}$$

As can be seen in (13), $S_i$ is the sum of all minimal principals of order $i$ of the partial transpose of the density matrix. The polynomial (12) has four roots and, if the quantum state is entangled, then one of them is negative. The absolute value of this root is the negativity given in (6). Hence, the entanglement measure can be stated as

$$N_e = -\min\{0, \lambda_{\min}\} \tag{14}$$

where $\lambda_{\min}$ is the minimal eigenvalue of (12).

**4. Conclusions**

We have presented a separability criterion and an entanglement measure based on the minimal principals of the partial transpose of the density matrix. The entanglement measure is the same negativity proposed by Vidal and Werner, but calculated using the minimal principals.